\documentclass[showpacs,aps,floatfix,prb,reprint,superscriptaddress]{revtex4-1}
\usepackage{xfrac}
\usepackage{amssymb}
\usepackage{braket}
\usepackage{graphicx}
\usepackage{verbatim}
\usepackage{etoolbox}\usepackage{amsfonts} %maths
\usepackage{amsmath} %maths
\usepackage[utf8]{inputenc} %useful to type directly diacritic characters
\usepackage{mathtools}
\usepackage{soul,xcolor,colortbl}
\usepackage{hyperref} \hypersetup{colorlinks=true,citecolor=blue,linkcolor=blue,urlcolor=blue} % HYPERLINKS
\usepackage{bm} %revtex4.1 for bold symbols
\usepackage{array}
\newcolumntype{C}[1]{>{\centering\arraybackslash}p{#1}}\usepackage{soul}
\definecolor{Gray}{gray}{0.85}

\usepackage{color} % STEFANO
\definecolor{Gray}{gray}{0.9}
\definecolor{LightCyan}{rgb}{0.88,1,1}

\begin{document}
\title{\Large \textbf{First principles thermodynamical modeling of the binodal and spinodal curves in lead chalcogenides}}

\author{Demet Usanmaz}
\affiliation{Department of Mechanical Engineering and Materials Science, Duke University, Durham, North Carolina 27708, USA}
\affiliation{Center for Materials Genomics, Duke University, Durham, NC 27708, USA}
\author{Pinku Nath}
\affiliation{Department of Mechanical Engineering and Materials Science, Duke University, Durham, North Carolina 27708, USA}
\affiliation{Center for Materials Genomics, Duke University, Durham, NC 27708, USA}
\author{Jose J. Plata}
\affiliation{Department of Mechanical Engineering and Materials Science, Duke University, Durham, North Carolina 27708, USA}
\affiliation{Center for Materials Genomics, Duke University, Durham, NC 27708, USA}
\author{Gus L. W. Hart}
\affiliation{Department of Physics and Astronomy, Brigham Young University, Provo, Utah 84602, USA}
\affiliation{Center for Materials Genomics, Duke University, Durham, NC 27708, USA}
\author{Ichiro Takeuchi}
\affiliation{Center for Nanophysics and Advanced Materials, University of Maryland, College Park, Maryland 20742, USA}
\affiliation{Department of Materials Science and Engineering, University of Maryland, College Park, Maryland 20742, USA}
\affiliation{Center for Materials Genomics, Duke University, Durham, NC 27708, USA}
\author{Marco Buongiorno Nardelli}
\affiliation{Department of Physics and Department of Chemistry, University of North Texas, Denton TX, USA}
\affiliation{Center for Materials Genomics, Duke University, Durham, NC 27708, USA}
\author{Marco Fornari}
\affiliation{Department of Physics, Central Michigan University, Mount Pleasant, MI 48858, USA}
\affiliation{Center for Materials Genomics, Duke University, Durham, NC 27708, USA}
\author{Stefano Curtarolo}
\affiliation{Materials Science, Electrical Engineering, Physics and Chemistry, Duke University, Durham NC, 27708}
\affiliation{Center for Materials Genomics, Duke University, Durham, NC 27708, USA}
\email{stefano@duke.edu}

\date{\today}

\begin{abstract}
High-throughput {\it ab-initio} calculations, cluster expansion techniques and thermodynamic modeling have been synergistically combined
to characterize the binodal and the spinodal decompositions features in the pseudo-binary lead chalcogenides PbSe-PbTe, PbS-PbTe,  and PbS-PbSe. 
While our results agree with the available experimental data, our consolute temperatures substantially improve with respect to previous computational modeling. 
The computed phase diagrams corroborate that the formation of spinodal nanostructures causes low thermal conductivities in these alloys. 
The presented approach, making a rational use of online quantum repositories, can be extended to study thermodynamical and kinetic properties of materials of technological interest.
\end{abstract}

%%%%%%%%%%%%%%%%%%%%%%%%%%%%%%%%%%%%%%%%%%%%%%%%%%%%%%%%%%%%%%%%%%%%%%%%%%%%%%%%%%%%%%%%%%%%%%%%%%%%%%%%%%

\maketitle

\section{Introduction}
For decades, the physical properties of lead chalcogenides have generated substantial interest in a number of fields,
in particular for applications in semiconductor technology.~\cite{Khokhlov-2003}
PbS, PbSe, and PbTe have distinct structural and electronic properties compared to 
III-V and II-VI compounds. These include high carrier mobilities, narrow band gaps with
negative pressure coefficients, high dielectric constants, and a positive temperature coefficient.~\cite{Adachi-2005,Moss-2013,Ravich-2013}
In addition, PbS, PbSe, and PbTe were predicted to be weak topological insulators, with a band inversion observed at the {\it N} point of the distorted
body-centered tetragonal Brilllouin zone.\cite{curtarolo:art77} 
These important and varied properties have allowed lead chalcogenides 
to be used extensively in optoelectronic devices such as lasers and detectors, thermophotovoltaic energy converters, and thermoelectric materials.~\cite{Alivisatos_science_1996, Khokhlov-2003, dalven_1973,Pei201240,Johari20125449,Wang20111366}

As thermoelectric materials, lead chalcogenides may exhibit electrical conductivities, ${\sigma}$, in excess of {\mbox {$2$-$4\cdot10^{-4}{\mathrm{\Omega^{-1}cm^{-1}}}$}}, 
thermopowers, $S$, around 150 ${\mathrm {\mu VK^{-1}}}$, 
and thermal conductivities, $\kappa$, on the order of $1$-$2~{\mathrm {Wm^{-1}K^{-1}}}$. This leads to
figures of merit $ZT={\sigma}S^2/\kappa$ larger than 1 at high temperatures, $T$.  
Such outstanding performances are due on the details of the electronic structure,~\cite{Heremans_science_2008,Parker201434,Parker2010,Parker2012,Mecholsky-prb-2014}
and on the ability to dramatically reduce the thermal conductivity with alloying and nanostructuring.~\cite{He-jacs-2010,Chena-pnsmi-2012,Harman-jeml-2005,Zhao-jacs-2012}
While pure lead chalcogenides are attractive on their own, their alloys are even more interesting. 
The appeal arises from their mechanical and electronic tunability, which can be optimized for specific technological needs.~\cite{zaoui_mcp_2009, naeemullaha_cms_2014, yamini_pccp_2014} 
For example, the PbTe$_{1-x}$Se$_x$ pseudo-binary system 
has a higher $ZT$ value than its corresponding binary forms.~\cite{Qzhang_jacs_2012, Pei_nature_2011}
Thallium doping in PbTe causes changes in the electronic density of states, increasing the $ZT$ value to 1.5 at 773~K.~\cite{Heremans_science_2008} Similarly, a  $ZT$ of 1.3 at 850~K was reported for aluminum doped PbSe.~\cite{Zhang_ees_2012}
 Furthermore, Pb$_{9.6}$Sb$_{0.2}$Te$_{10-x}$Sb$_x$ is known to exhibit lower thermal conductivity and a higher $ZT$ than PbSe$_{1-x}$Te$_x$.~\cite{poudeu_jacs_2006}
This is also true for nanostructured (Pb$_{0.95}$Sn$_{0.05}$Te)$_{0.92}$(PbS)$_{0.08}$, as the low thermal conductivity leads to a $ZT=$ 1.5 at 642~K.~\cite{Androulakis_jacs_2015}
The properties of this group of pseudo-binaries depend greatly on the atomic details of the material's morphology. 
This can be partially understood in terms of thermodynamical features. 
The excellent thermoelectric performances, for example, were tentatively ascribed to the limited miscibility of the components. This gives rise to structural inhomogeneities that lower the thermal conductivity without damaging the electronic transport.~\cite{Zhao-jacs-2012} Such control over the morphology could be also used to optimize functionalities associated with topological effects.
 
In this work we study the phase diagram of lead chalcogenide pseudo-binaries. 
We predict quantitatively the boundary of the solid solution (the binodal curve that defines the region of miscibility), as well as the spinodal region. 
These features are key for rationalizing and honing synthesis and characterization of optimized systems. 
To the best of our knowledge, our study is the first to completely and accurately report such characterization. 
The phase diagram is essential for properly establishing manufacturing processes. There has been only one previous attempt to model phase diagrams of lead chalcogenides using thermodynamic modeling (TM),~\cite{boukhris_ps_2011} which predicted 
consolute temperature ({\it T$_c$}) values far from those reported in experimental studies.~\cite{Liu_MineMag_1994} 
The disagreement was attributed to the difficulty of an exhaustive exploration of the different configurations for each composition. 
Here, we built upon the synergy between cluster expansion (CE) techniques, high-throughput (HT) \textit{ab initio} calculations,~\cite{curtarolo:art49,curtarolo:art53,curtarolo:art56} and thermodynamical modeling to find acceptable agreement between our Monte Carlo simulations (MC) and the available experimental results.

%%%%%%%%%%%%%%%%%%%%%%%%%%%%%%%%%%%%%%%%%%%%%%%%%%%%%%%%%%%%%%%%%%%%%%%%%%%%%%%%%%%%%%%%%%%%%%%%%%%%%%%%%%

\section{\label{sec:met}Methodology}

\subsection{Thermodynamic Modeling}
Pseudo-binary systems are represented by the formula 
{\it (A$_{x_{A}}$B$_{x_ {B}}$)$_a$C$_c$} (or {\it A$_x$B$_{1-x}$C}) with mole fractions {\it $x_{A}$} and {\it $x_{B}$} of elements {\it A} and {\it B} respectively, related by {\it $x_{A}$+$x_{B}$}=1. The small letters {\it a} and {\it c} represent number of sites per formula. \cite{saunders_1998,hillert_2008} The Gibbs energy of such iso-structural pseudo-binary systems can be written as:~\cite{saunders_1998,hillert_2008}
\begin{equation}
\label{gibbs}
\begin{split}
{\it G}_{(\it{A,B})_{a}\it{C}_{c}} = x_{\it A}{\it G}_{\it A_{a}\it C_{c}}+x_{\it B}G_{\it B_{a}\it C_{c}} 
\\ +k_{\it B}{\it T}(x_{\it A}\it ln( x_{\it A})+x_{\it B}\it ln (x_{\it B}))+x_{\it A}x_{\it B}\it L_{\it {A,B:C}},
\end{split}
\end{equation}
where {\it G$_{A_aC_c}$} and {\it G$_{B_aC_c}$} represent the Gibbs free energy of {\it A$_a$C$_c$} and {\it B$_a$C$_c$} materials. 
These two variables can be computed at any temperature by fitting available  experimental 
data~\cite{Barin} to the polynomial form~\cite{Zhang_ActaMat_2007} shown in Equation~(\ref{gibbs2}):
\begin{equation}
\label{gibbs2}
{\it G}(T) = {\it a} + {\it b} \it T + {\it c} \it T\it {ln T} +  {\it d}  \it T^{2} + {\it e} \it T^{-1} + {\it f} \it T^{3}.
\end{equation}
The third term in the Equation~(\ref{gibbs}) is the entropy of mixing, and the last term is the excess energy of mixing that 
represents the non-ideality of the system. In contrast to the entropic term, the excess energy
parameter can take negative or positive values. If {\it L$_{A,B:C}$} is negative, it indicates that the 
system tends to create a solid solution. A positive value of {\it L$_{A,B:C}$} indicates a repulsive interaction between phases, penalizing formation of intermediate alloys.
To find 
the excess mixing energy, the composition-dependent interaction parameter {\it L$_{A,B:C}$} can be calculated 
with Equation~(\ref{DHxaxbL}):
\begin{equation}
\Delta \it H=x_{\it A}x_{\it B}\it L_{\it{A,B:C}}.
\label{DHxaxbL}
\end{equation}

The enthalpy of formation, $\Delta${\it H}, is defined as:
\begin{equation}
\Delta \it H=\it{E}_{\it{(A,B)}_{a}\it C_{c}}- x_{A}\it{E}_{\it A_{a}\it C_{c}} - x_{\it B}\it{E}_{\it B_{a}\it C_{c}},
\end{equation}
where {\it E$_{{(AB)}_{a}C_{c}}$}, {\it E$_{A_{a}C_{c}}$}, and {\it E$_{B_{a}C_{c}}$} are the total energies of compounds 
{\it (AB)$_{a}$C$_{c}$}, {\it A$_{a}$C$_{c}$} and {\it B$_{a}$C$_{c}$}, respectively. These energies can be found from the fully relaxed structures using density functional theory (DFT).

A combination of high throughput {\it ab initio} calculations and thermodynamic 
modelling are used to predict the interaction parameter, {\it L$(x)$}.~\cite{Zhang_ActaMat_2007, Sheng_ActaMat_2011, SHSheng_ActaMat_2011} The result is a zero temperature approximation of the actual value. 
The most common method to describe the composition dependent interaction parameter is the Redlich-Kister equation, 
~\cite{Redlich_IndsEngChem_1948,saunders_1998,hillert_2008} where the interaction parameter is written in a polynomial form:
\begin{equation}
\it{L}(x)= \sum_{i=0}^{n} \it L_{i}(x_{\it A}-x_{\it B})^n.
\label{interaction_parameter}
\end{equation}
We fit this polynomial to the formation enthalpy data obtained from DFT calculations. We checked that an 
$n=2$ polynomial is enough to obtain a good fit to the data, so that only {\it L$_0$}, {\it L$_1$}, and {\it L$_2$} need to be determined:
\begin{equation}
\it{L}(x)\simeq \it L_0+x\it L_1+x^2\it L_2
\label{polynomial}
\end{equation}
to compute the interaction parameter.

The Gibbs free energy is composition and temperature dependent. 
The main computational challenge lies in characterizing many configurations for many compositions.
Some authors have attempted to tackle the issue, by generating, few configurations and/or few 
compositions. ~\cite{boukhris_ps_2011, Zhang_ActaMat_2007, Sheng_ActaMat_2011, SHSheng_ActaMat_2011}
Here, we have chosen to challenge the issue more drastically,
by reling on the advantages of a hybrid cluster expansion - high throughput approach \cite{curtarolo:art81} featuring:
exhaustive exploration of different configurations for different compositions (CE),
minimization of computational cost by reducing the number of the {\it ab-initio} calculations (CE),
analysis of many compositions as typical of HT methods \cite{curtarolo:art81,aflowBZ},
and
rational use of online repositories ({\sf AFLOWLIB.org}) \cite{aflowlibPAPER}.

%-------------------------------------------------------------------------------------------

\subsection{ Cluster expansion} \label{clusterexpansion}

In the cluster expansion technique, the configurational energy  of an alloy as written as a sum of many-body occupation 
variables $\left\{\sigma\right\}$:~\cite{deFontaine_ssp_1994}
\begin{equation}
{\it E(\sigma)}= {\it J_0} + \sum_i {\it J_i \sigma_i} + \sum_{ij} {\it J_{ij}\sigma_{ij}} + \dots,
\end{equation}
where {\it $J_0$}, {\it $J_i$}, {\it $J_{ij}$}, \textit{etc}. are known as effective cluster interactions and must be determined. 

The above equation can be rewritten into symmetrically distinct sets of clusters, $\alpha$:
\begin{equation}\label{CEsymmetricEQ}
\it{E}(\sigma)=\sum\limits_{\alpha} {\it m}_{\alpha}{\it J}_{\alpha} \langle \prod\limits_{i{\in}{\alpha}} {\sigma}_i\rangle,
\end{equation}
 where $m_\alpha$ represents symmetrically equivalent clusters $\alpha$ in a given reference volume.~\cite{burton_cg_2006}
The $J_\alpha$ parameters are obtained by fitting a relatively small number of DFT calculated energies.

The reliability of the predicted energy may be determined using the cross-validation score CV:
\begin{equation}\label{CECV}
\it{(CV)}^2=\frac{1}{N}\sum\limits_{S=1}^{N} (\it{E}_{S}-\hat{\it{E}}_{S})^2,
\end{equation}
 where E$_{S}$ represents the calculated energies from DFT and \^{E$_{S}$} is the predicted energies from CE.
 
The enumeration of configurations, calculation of the effective interaction parameters,
determination of ground state structures, and prediction of more structures was performed with
the Alloy Theoretic Automated Toolkit (ATAT).~\cite{Walle_calphad_2002} 
Calculated phase diagrams were obtained with Monte Carlo (MC) simulations performed with $phb$ code. 
The algorithm automatically follows a given phase boundary and is provided by the ATAT 
package.~\cite{Walle_calphad_2002,Walle_jpe_2002,Walle_msmse_2002}
%------------------------------------------------------------------------------------------------------

\subsection{High-throughput {\bf \it ab initio} calculations} \label{HTDFT}

All DFT calculations were carried out by using the 
Automatic-Flow for Materials Discovery ({\small AFLOW}) ~\cite{curtarolo:art65,aflowlibPAPER,curtarolo:art92} 
and DFT Vienna {\it ab initio} simulation program  ({\small VASP}). \cite{kresse_vasp} Calculations were performed using 
{\small AFLOW} standards.~\cite{calderon_2015} We use the projector augmented wave (PAW) pseudopotentials~\cite{PAW} 
and the exchange and correlation functionals parametrized by the generalized gradient approximation proposed 
by Perdew-Burke-Ernzerhof.~\cite{PBE} All calculations use a high energy-cutoff, which is 40$\%$ larger 
than the maximum cutoff of all pseudopotentials used. 
Reciprocal space integration was performed using 8000 more k-points than the number of atoms.
{Spin-orbit  coupling  was  not  treated
in  the  calculations due to its minimal influence in $\Delta${\it H} (smaller than 1.5 meV/atom).}
Structures were
fully relaxed (cell volume and ionic positions) such that the energy difference between two consecutive ionic steps was smaller than $10^{-4}$ eV.

PbS, PbSe and PbTe crystallize in the NaCl structure and belong to the $Fm\overline{3}m$ space group (\# 225). 
Space and point group symmetries of intermediate composition structures were determined using {\small AFLOW}.

%%%%%%%%%%%%%%%%%%%%%%%%%%%%%%%%%%%%%%%%%%%%%%%%%%%%%%%%%%%%%%%%%%%%%%%%%%%%%%%%%%%%%%%%%%%%%%%%%%%%%%%%%%

\section{\label{sec:results}Results and Discussion}

\subsection{The PbSe$_{1-x}$Te$_x$ alloy}
\begin{figure}[]%
\includegraphics*[width=8.5cm,clip=true]{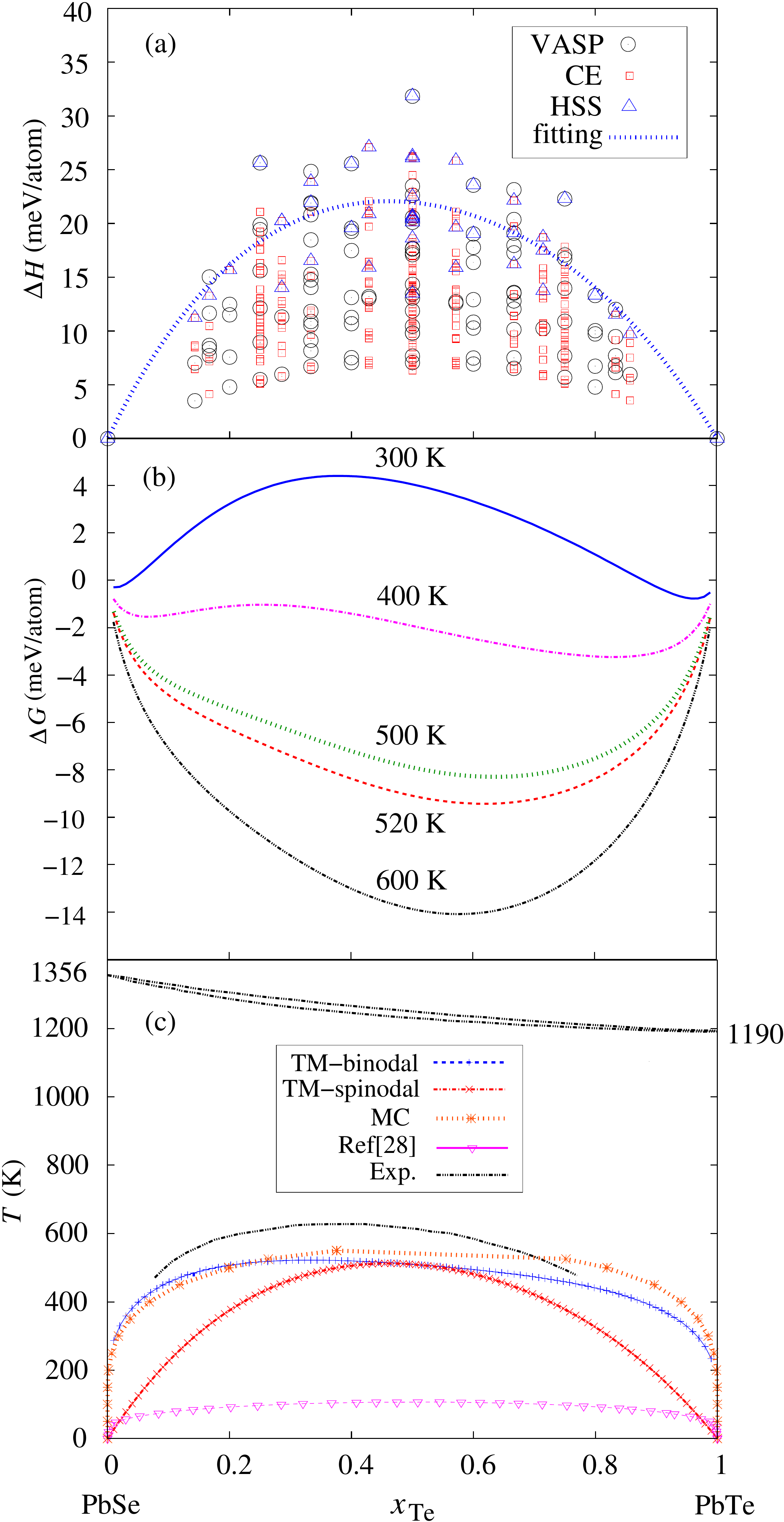}
\caption{ (a) Formation enthalpies of the PbSe-PbTe structures using DFT calculations ($\bigcirc$) 
and CE technique ({\color{red}{$\Box$}}). Highly symmetryic structures are represented by {\color{blue}{$\bigtriangleup$}} and the fitting of these points to obtain the interaction parameter is plotted with a blue dashed line. (b) {$\Delta$}$G(x)$ diagram at various temperatures. (c) Binodal and spinodal curves from TM ({\color{blue}{$+$}} and {\color{red}{$+$}}), and MC simulations ({\color{orange}{$\ast$}}), compared with experimental data \cite{Liu_MineMag_1994}  ($-\cdot\cdot\cdot$) and the previous theoretical model ({\color{magenta}{$\bigtriangledown$}}) from Boukhris {\emph et al.}~\cite{boukhris_ps_2011}}
\label{PbSeTe}
\end{figure}

In agreement with experimental data, we found that PbSe and PbTe are immiscible systems at 0~K. This is shown in Figure.~\ref{PbSeTe}(a), 
where formation enthalpies are positive for all compositions ($0<x<1$). The CE predicted 
energies are in excellent agreement with DFT calculated structures, with a cross validation 
score of 4$\cdot10^{-4}$. 
Our quantitative results confirm the  Hume-Rothery rules.~\cite{cottrell1967}
These rules qualitatively predict the miscibility of two metals based on four properties: atomic radius, crystal lattice, valence and electronegativity.
Amongst the chalcogens, the atomic radius changes from 1.04 {\AA} in S, to 1.17 {\AA} in Se, to 1.37 {\AA} in Te. 
These size variations create a mismatch between the lattice parameters of PbTe and PbSe, causing incoherence in the interface and phase decomposition, eventually. 
{$\Delta$}{\it G(x)} diagrams at different temperatures are plotted in Figure.~\ref{PbSeTe}(b). 
It can be seen the {$\Delta$}{\it G} function has two minima and a single maximum at low temperatures. At high temperatures close to $T_c$, it becomes convex, with one minimum.

The binodal curve $T_c(x)$ is defined by the horizontal tangent points of the Gibbs free energy, {\it G}. When $T<T_c$, the alloy starts decomposing. 
Additionally, our calculations let us determine the spinodal curve that discriminates metastable and unstable regions in the pseudo-alloy phase diagram.
The spinodal curve is the locus of the points where the second derivative of 
{\it G} is equal to 0:
\begin{equation}
\frac{\partial^{2}{\it G}_{(\it{AB})_{a}\it{C}_{c}}}{\partial x^{2}}=0.
\end{equation}

We computed first and  second derivatives of the Gibbs free energy within our thermodynamical model.
In order to obtain the $L_0$, $L_1$, $L_2$ that are  necessary to compute $L(x)$ at any composition, we use highly symmetric structures (HSS)~\cite{Ferreira_PRB_2006,Marques_PRB_2007} to fit Equation~(\ref{DHxaxbL}). 
HSS have a larger degeneracy and thus greater weight in the properties of the ensemble; particularly at high temperatures that are close to the spinodal decomposition. 
After computing the $L_n$ constants, $G_{(Se,Te)_{a}Pb_{c}}$ is obtained using Equation~(\ref{gibbs}). 
The results obtained from the thermodynamic model are compared with our MC results, previous 
theoretical predictions,~\cite{boukhris_ps_2011}  and experimental data.~\cite{Liu_MineMag_1994,Darrow_JMatSci_1969}

Our calculations encompass the entire range of concentrations, and reproduce the asymmetry observed experimentally in the binodal curve. 
In order to  quantitatively compare all data, 
we analyzed the consolute temperature or upper critical solution temperature ($T_c$) describing the lowest temperature at which both phases are miscible  at any composition. 
Experimentally PbSe$_{1-x}$Te$_x$ alloy presents, at  $x=0.4$, a $T_c$ 
closer to 623~K.~\cite{Liu_MineMag_1994,Darrow_JMatSci_1969}
This quantity is far 
from the value predicted by Boukhris \emph{et al.}~\cite{boukhris_ps_2011}, which is around 106~K.  
The consolute temperature predicted by our thermodynamic model at $~$520~K (16.5\% error) is around $x=0.34$ (see Figure.\ref{PbSeTe}(c)). 
Our results quantitatively improve the prediction of T$_c$ with of Boukhris {\emph et al}.~\cite{boukhris_ps_2011} 
Moreover, this value approaches the results obtained by much more expensive techniques  such as MC, in which we obtain a value close to 550~K (11.7\% error).
Discrepancies between MC and TM at larger Te (Se) concentrations arise from difficulties in converging MC calculations in the dilute limit of Se (Te).
The experimental miscibility gap presents a slight asymmetric form that is reproduced by  MC and our  TM. 
This asymmetry is observed in experiments, but was not seen in previous theoretical work.\cite{boukhris_ps_2011}. This phenomenon will be discussed in the next section.

In contrast to the binodal curve, the spinodal curve is quite symmetric. 
The combination of the symmetric spinodal and asymmetric binodal curves at the Se-rich region causes nucleation at higher temperatures. This information is very important for fine tuning synthesis protocols to obtain the desired morphologies. 

%Fig 1

\subsection{The PbS$_{1-x}$Te$_x$ alloy} 
The atomic radius of Te is 24\% larger than the S radius. Thus, the PbS-PbTe system follows the same trend as PbSe-PbTe, and they are immiscible at 0~K.

\begin{figure}[]%
\includegraphics*[width=8.5cm,clip=true]{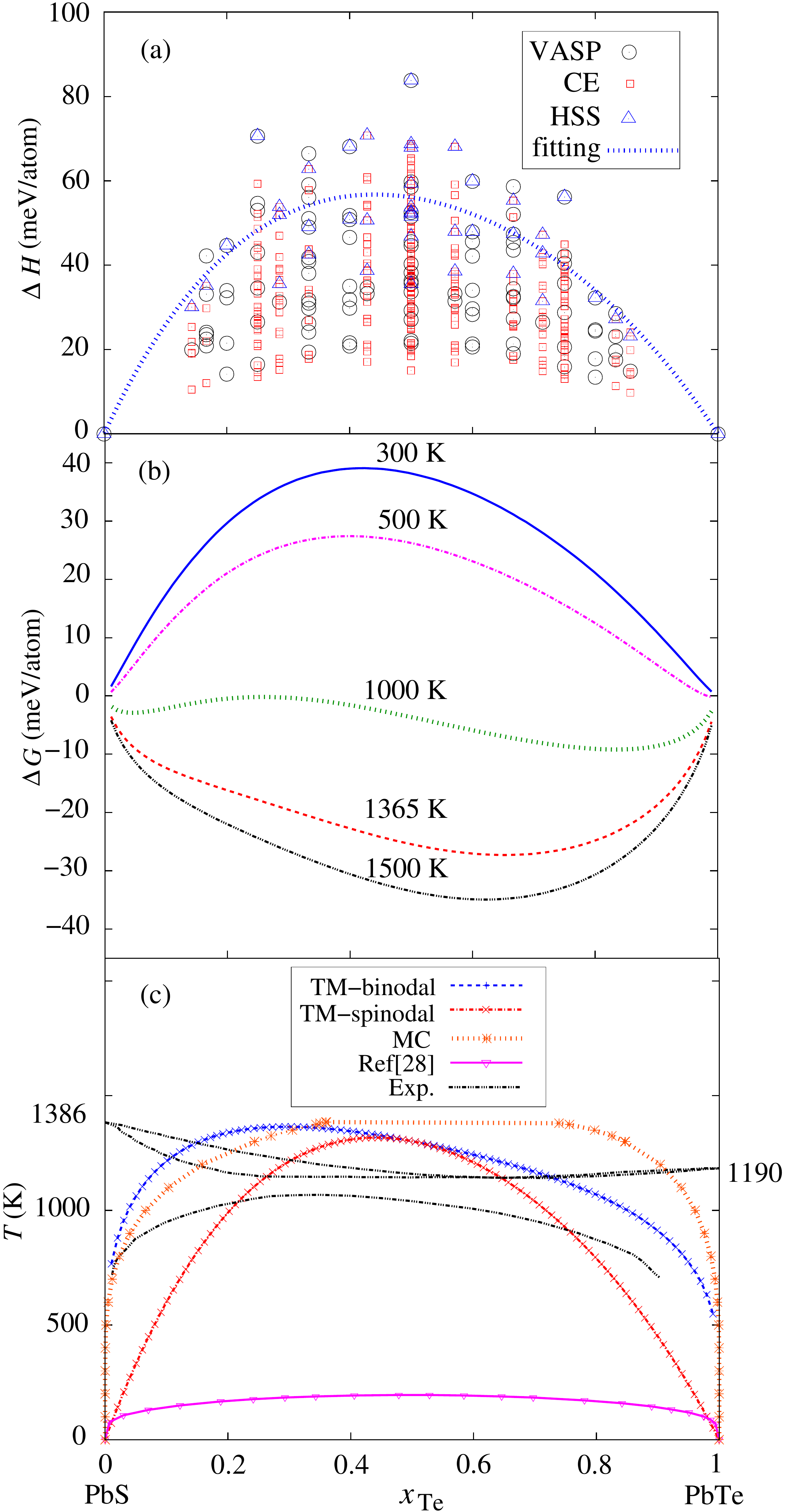}
\caption{(a) Formation enthalpies of the PbS-PbTe structures using DFT calculations ($\bigcirc$) 
and CE technique ({\color{red}{$\Box$}}). Highly symmetric structures are represented 
by {\color{blue}{$\bigtriangleup$}} and the fitting of these points to obtain the interaction parameter 
is plotted with a blue dashed line. (b) {$\Delta$}$G(x)$ diagram at various temperatures. (c) Binodal and spinodal curves from TM ({\color{blue}{$+$}} and {\color{red}{$+$}}), 
and MC simulations ({\color{orange}{$\ast$}}), compared with experimental data \cite{Liu_MineMag_1994}  ($-\cdot\cdot\cdot$)
and the previous theoretical model ({\color{magenta}{$\bigtriangledown$}}) from Boukhris {\emph et al.}~\cite{boukhris_ps_2011}}
\label{PbSTe}
\end{figure}

Our {$\Delta$}$H$ values are greater than 0 eV for the whole range of concentrations (see Figure.~\ref{PbSTe}(a)). 
CE predicted energies are again in agreement with DFT calculations, obtaining a CV score of about  3$\cdot 10^{-3}$. 
The fitting for $L(x)$ is also depicted in  Figure.~\ref{PbSTe}(a) using highly symmetric points.
{Similarly to PbSe-PbTe, the {$\Delta$}$G$ function changes  to a convex shape at high temperatures (Figure.~\ref{PbSTe}(b)).}

%Fig 2

The calculated phase diagram of the PbS$_{1-x}$Te$_x$ alloy is shown in Figure.~\ref{PbSTe}(c). 
Experimental results show again a slight asymmetry with a maximum around $x=0.3$. This is 
in agreement with our results, while MC simulations fail to show this
asymmetry. The predicted consolute temperatures for PbS-PbTe follow the same trend as 
PbSe-PbTe. Results published by Boukhris \emph{et al.}~\cite{boukhris_ps_2011} considerably underestimate the experimental value for $T_c$ (1083~K).
Our prediction of  the consolute temperature is slightly larger than 
experiments~\cite{Liu_MineMag_1994}; being 1385 and 1365~K using MC simulations and TM, respectively.  
{As seen from Figure.~\ref{PbSeTe}(c)  and Figure.~\ref{PbSTe}(c) PbSe-PbTe and PbS-PbTe systems show a very similar trend of a slightly asymmetric spinodal curve, and considerably asymmetric binodal curve. This trend shows that formation of the Te-rich alloy starts at lower temperatures than Se-rich compositions.}

\subsection{The PbS$_{1-x}$Se$_x$ alloy} 
\begin{figure}[]%
\includegraphics*[width=8.5cm,clip=true]{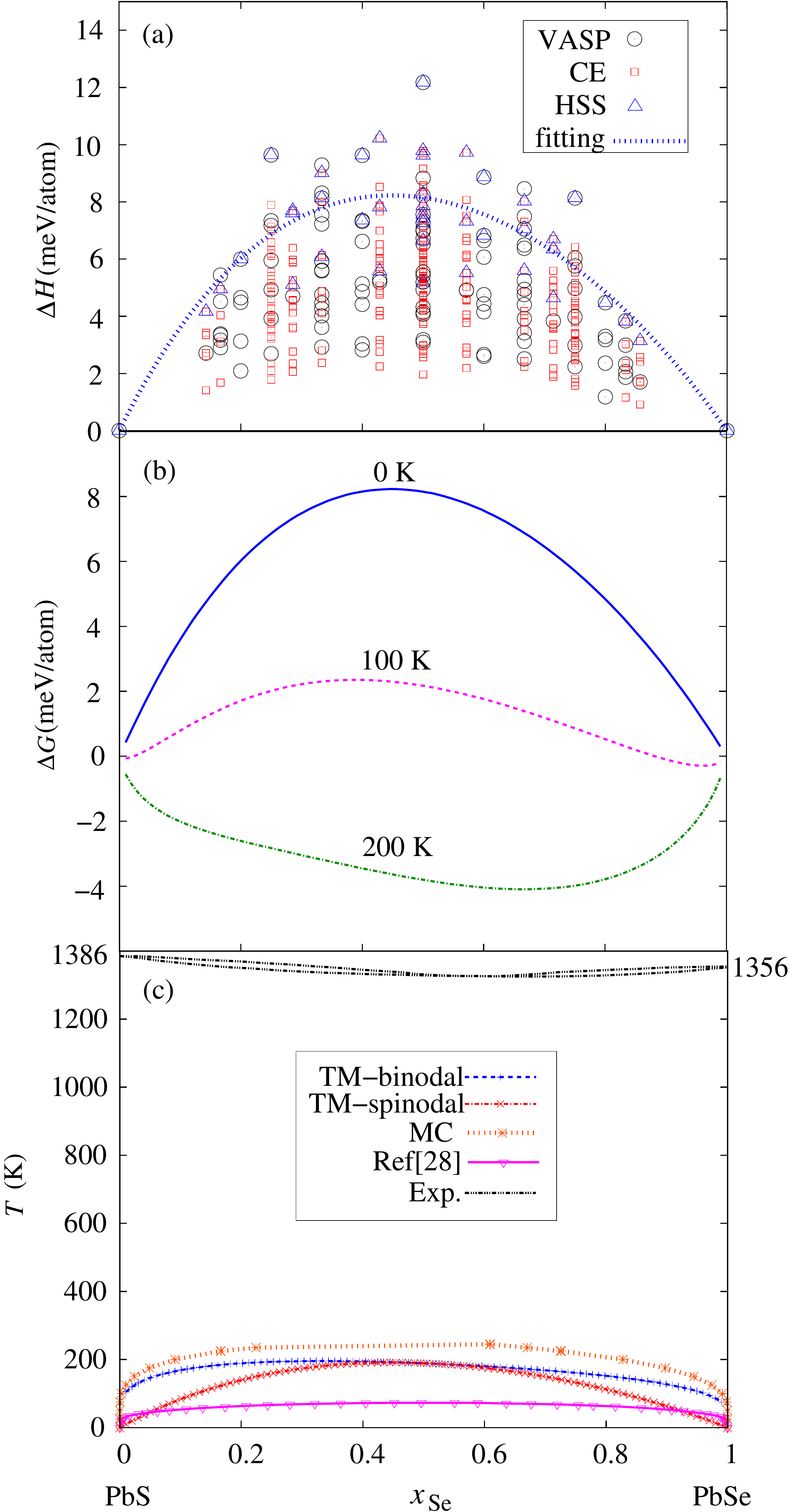}
\caption{ (a) Formation enthalpies of the PbS-PbSe structures using DFT calculations ($\bigcirc$) 
and CE technique ({\color{red}{$\Box$}}). Highly symmetric structures are represented 
by {\color{blue}{$\bigtriangleup$}} and the fitting of these points to obtain the interaction parameter 
is plotted with a blue dashed line. (b) {$\Delta$}$G(x)$ diagram at various temperatures. (c) Binodal and spinodal curves from TM ({\color{blue}{$+$}} and {\color{red}{$+$}}), 
and MC simulations ({\color{orange}{$\ast$}}), compared with experimental data \cite{Liu_MineMag_1994}  ($-\cdot\cdot\cdot$)
and the previous theoretical model ({\color{magenta}{$\bigtriangledown$}}) from Boukhris {\emph et al.}~\cite{boukhris_ps_2011}}
\label{PbSSe}
\end{figure}

Our methodology was also applied to the PbS-PbSe system. All positive energies in Figure.~\ref{PbSSe} 
indicate that PbS and PbSe systems are not miscible at 0~K. As far as we know, there are no 
experimental data available below 573~K~\cite{Darrow_JMatSci_1969} for this system. However, 
it has been shown that MC simulations predict the $T_c$ for different systems quite well, and can describe the miscibility gap.~\cite{burton_cg_2006, burton_jap_2011} For this alloy, 
thermodynamic modelling predicts a $T_c$ slightly below 200~K and MC predicts a $T_c$ slightly lower 
than 250~K.

%Fig 3

\subsection{General considerations} 
\begin{figure}%
\includegraphics*[width=8cm,clip=true]{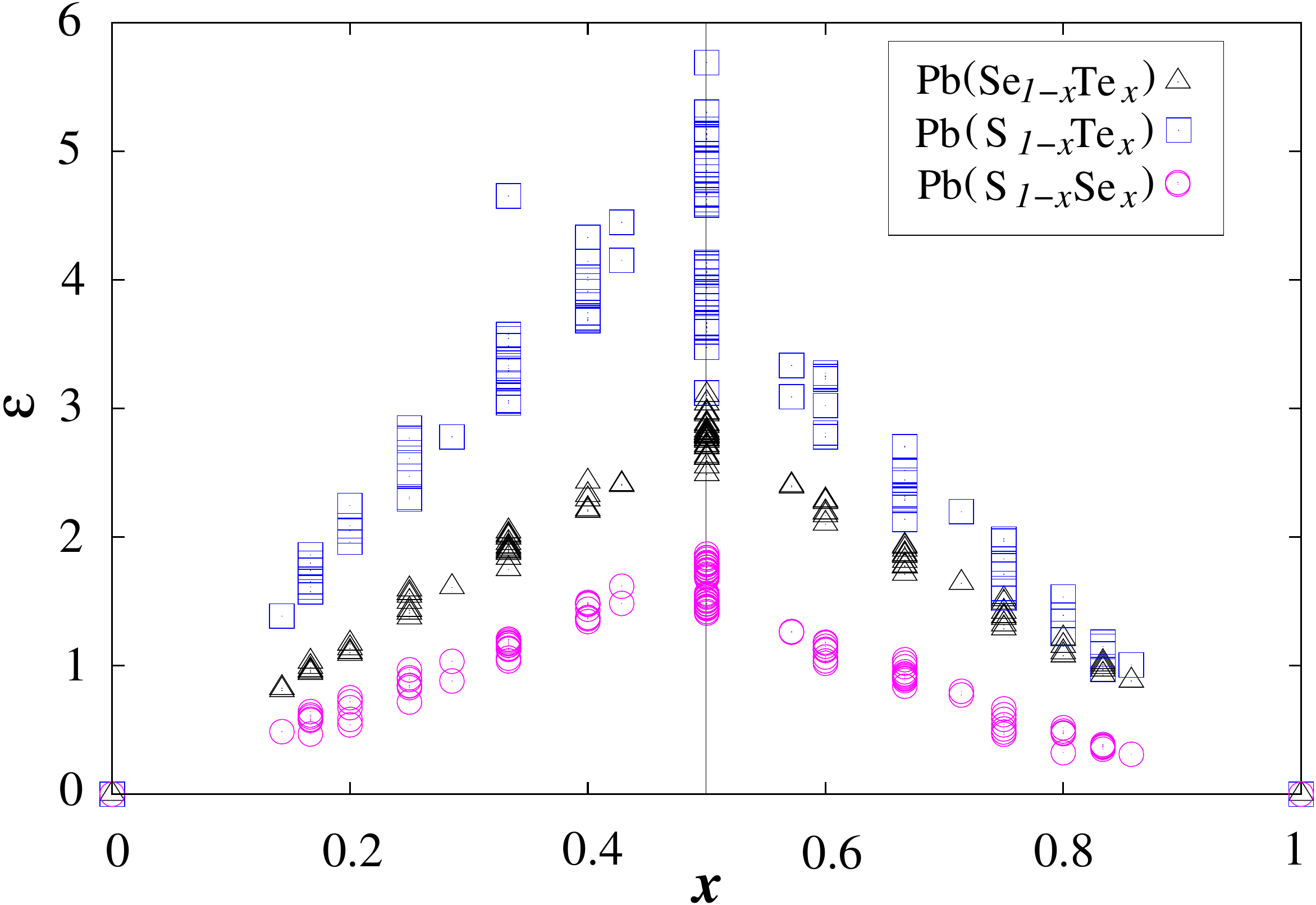}
\caption{Lattice mismatch for lead chalcogenides alloys. For each ternary system we consider $0$ mismatch when $x=0$.}
\label{volume}
\end{figure}

{The lattice mismatch between the two solids (PbS, PbSe, or PbTe) is a good descriptor to
analyze the trends observed experimentally (see Figure.~\ref{volume}). Lattice mismatch, $\epsilon$, is defined as:
\begin{equation}
\epsilon = {\frac{\it (a_{(A,B)_{a}C_{c}}-a_{\mathrm {solvent}})}{a_{\mathrm {solvent}}} {\times}100},
\end{equation}
where a$_{(A,B)_{a}C_{c}}$ denotes the lattice constant of intermediate alloys and a$_{\mathrm {solvent}}$ is the lattice constant of the most abundant binary alloy.} There is a correlation between the lattice mismatch of the alloy, {$\Delta$}$H$, and the consolute temperature. The higher the mismatch, the higher {$\Delta$}$H$ becomes and thus, a higher $T_c$ is obtained. For instance, the larger mismatch corresponds to the PbS-PbTe system, which presents a maximum enthalpy of formation at $x=0.5$, 
with 80 meV/atom and a $T_c$ of 1083~K. For the PbSe-PbTe system, the maximum $\Delta${\it H} is around 22 meV/atom and the 
consolute temperature is 623~K. Following this trend, PbS-PbSe system presents a smaller mismatch and a 
smaller {$\Delta$}$H$ 8 meV/atom. Thus, a consolute temperature smaller than 623~K is expected. 
If we approximate this correlation to a linear function, for a lattice mismatch around 3\% we get a $T_c$ close to 270~K; which is in agreement with MC and our thermodynamic model results.

%Fig 4

{Mismatch between lattices can be also used to explain the asymmetry of the binodal curves. 
We can define the asymmetry of the curve as the ratio between the decomposition temperature 
of two points equidistant to $x=0.5$. We have chosen 0.2 and 0.8 to define our asymmetry descriptor, $\epsilon_T$:
\begin{equation}
\epsilon_T= \frac{T(x=0.8)}{T(x=0.2)}.
\label{aT}
\end{equation}
Using this definition, we can assume that a perfectly symmetric spinodal curve has $\epsilon_T$=1. 

As discussed above, mismatch between lattices is directly related to the magnitude and size of the spinodal curve. Mismatch is the driving force in the three systems we are studying. Thus, we propose a second asymmetry descriptor, $\epsilon_r$, based on the ratio between the lattice mismatch at two points equidistant to $x=0.5$:
\begin{equation}
\epsilon_r = \frac{\epsilon(x=0.8)}{\epsilon(x=0.2)}.
\label{ae}
\end{equation}
The asymmetry descriptor values for the three systems are shown in Table~\ref{tab:asym}. $\epsilon_T$ shows PbS-PbSe as the most asymmetric spinodal curve, then PbS-PbTe, and finally the PbSe-PbTe system. This trend is exactly the same for $\epsilon_r$, emphasizing the importance of the lattice strain in the  of these systems.
   \begin{table}[h]
     \centering
     \begin{tabular}[b]{c  c  c  c}
       \hline
                & PbSe-PbTe & PbS-PbTe & PbS-PbSe  \\
       \hline
        $\epsilon_T$             &  0.86  &    0.80     &  0.80          \\
        $\epsilon_r$      &  0.84  &    0.59     &  0.49          \\
       \hline
     \end{tabular}
     \caption{Asymmetric descriptors for spinodal curves for lead chacogenides}
     \label{tab:asym}
   \end{table}
}

\section{\label{sec:conc}Conclusions}
A hybrid approach, comprising high-throughput {\it ab-initio} and cluster-expansion techniques
is used to create a thermodynamic model for calculating binodal and spinodal decompositions 
in pseudo binary lead chalcogenides (PbSe-PbTe and PbS-PbTe).
The model overcomes the limitations of previous theoretical studies, where too few compositions and/or configurations were taken into account. 
The obtained thermodynamical features are very close to the experimentally data, when available. 
We also capture the asymmetry of the binodal curve, experimentally observed and previously computationally unresolved.
Additionally, phase diagrams for systems without experimental characterization, such as the PbS-PbSe alloy, are suggested.
The results have been valitaded by using MC simulations, and lattice mismatch between the binary solids descriptors. 

Overall the approach is suitable for the high-throughput characterization of miscibility gaps, spinodal and other decomposition phenomena.

\section{\label{sec:ack}Acknowledgment}
We thank Dr. Allison Sterling and Dr. Cormac Toher for various technical discussions. We would like to acknowledge support by the by DOD-ONR (N00014-13-1-0635, N00014-11- 1-0136, N00014-09-1-0921). The AFLOWLIB consortium would like to acknowledge the Duke University Center for Materials Genomics and the CRAY corporation for computational support.

%------------------------------------------------------------------------------------------------------------

\bibliographystyle{PhysRevwithTitles_noDOI_v1b} %EOPAPER
\bibliography{xstefano4} %EOPAPER

\newcommand{\Ozolins}{Ozoli\c{n}\v{s}}
\begin{thebibliography}{10}
\expandafter\ifx\csname urlstyle\endcsname\relax
  \providecommand{\doi}[1]{doi:\discretionary{}{}{}#1}\else
  \providecommand{\doi}{doi:\discretionary{}{}{}\begingroup
  \urlstyle{rm}\Url}\fi
\providecommand{\selectlanguage}[1]{\relax}
\providecommand{\bibAnnoteFile}[1]{%
  \IfFileExists{#1}{\begin{quotation}\noindent\textsc{Key:} #1\\
  \textsc{Annotation:}\ \input{#1}\end{quotation}}{}}
\providecommand{\bibAnnote}[2]{%
  \begin{quotation}\noindent\textsc{Key:} #1\\
  \textsc{Annotation:}\ #2\end{quotation}}

\bibitem{Khokhlov-2003}
D.~Khokhlov, \emph{Lead Chalcogenides: Physics and Applications},
  Optoelectronic properties of semiconductors and superlattices (Taylor \&
  Francis, Great Britain, 2002).
\bibAnnoteFile{Khokhlov-2003}

\bibitem{Adachi-2005}
S.~Adachi, \emph{Properties of Group-IV, III-V and II-VI Semiconductors} (John
  Wiley \& Sons, England, 2005).
\bibAnnoteFile{Adachi-2005}

\bibitem{Moss-2013}
T.~Moss, G.~Burrell, and B.~Ellis, \emph{Semiconductor Opto-Electronics}
  (Butterworth-Heinemann, 2013).
\bibAnnoteFile{Moss-2013}

\bibitem{Ravich-2013}
Y.~I. Ravich, \emph{Semiconducting Lead Chalcogenides} (Springer Science
  Business Media, New York, 2013).
\bibAnnoteFile{Ravich-2013}

\bibitem{curtarolo:art77}
K.~Yang, W.~Setyawan, S.~Wang, M.~{Buongiorno~Nardelli}, and S.~Curtarolo,
  \emph{A search model for topological insulators with high-throughput
  robustness descriptors}, Nat.\ Mater. \textbf{11}, 614--619 (2012).
\bibAnnoteFile{curtarolo:art77}

\bibitem{Alivisatos_science_1996}
A.~P. Alivisatos, \emph{Semiconductor Clusters, Nanocrystals, and Quantum
  Dots}, Science \textbf{271}, 933--937 (1996).
\bibAnnoteFile{Alivisatos_science_1996}

\bibitem{dalven_1973}
R.~Dalven, H.~Ehrenreich, F.~Seitz, and D.~Turnbull, \emph{Solid State Physics}
  (Academic. Press, New York, 1973).
\bibAnnoteFile{dalven_1973}

\bibitem{Pei201240}
Y.-L. Pei and Y.~Liu, \emph{Electrical and thermal transport properties of
  Pb-based chalcogenides: PbTe, PbSe, and PbS}, J. Alloy. Comp. \textbf{514},
  40--44 (2012).
\bibAnnoteFile{Pei201240}

\bibitem{Johari20125449}
P.~Johari and V.~B. Shenoy, \emph{Tuning the electronic properties of
  semiconducting transition metal dichalcogenides by applying mechanical
  strains}, ACS Nano \textbf{6}, 5449--5456 (2012).
\bibAnnoteFile{Johari20125449}

\bibitem{Wang20111366}
H.~Wang, Y.~Pei, A.~D. Lalonde, and G.~J. Snyder, \emph{Heavily doped p-type
  PbSe with high thermoelectric performance: An alternative for PbTe}, Adv.
  Mater. \textbf{23}, 1366--1370 (2011).
\bibAnnoteFile{Wang20111366}

\bibitem{Heremans_science_2008}
J.~P. Heremans, V.~Jovovic, E.~S. Toberer, A.~Saramat, K.~Kurosaki,
  A.~Charoenphakdee, S.~Yamanaka, and G.~J. Snyder, \emph{Enhancement of
  Thermoelectric Efficiency in PbTe by Distortion of the Electronic Density of
  States}, Science \textbf{321}, 554--557 (2008).
\bibAnnoteFile{Heremans_science_2008}

\bibitem{Parker201434}
D.~Parker and D.~Singh, \emph{High temperature thermoelectric properties of
  rock-salt structure PbS}, Solid State Commun. \textbf{182}, 34--37 (2014).
\bibAnnoteFile{Parker201434}

\bibitem{Parker2010}
D.~Parker and D.~Singh, \emph{High-temperature thermoelectric performance of
  heavily doped PbSe}, Phys.\ Rev.\ B \textbf{82}, 035204 (2010).
\bibAnnoteFile{Parker2010}

\bibitem{Parker2012}
D.~Parker, D.~Singh, Q.~Zhang, and Z.~Ren, \emph{Thermoelectric properties of
  n-type PbSe revisited}, J.\ Appl.\ Phys. \textbf{111}, 123701 (2012).
\bibAnnoteFile{Parker2012}

\bibitem{Mecholsky-prb-2014}
N.~A. Mecholsky, L.~Resca, I.~L. Pegg, and M.~Fornari, \emph{Theory of band
  warping and its effects on thermoelectronic transport properties}, Phys. Rev.
  B \textbf{89}, 155131 (2014).
\bibAnnoteFile{Mecholsky-prb-2014}

\bibitem{He-jacs-2010}
J.~He, J.~R. Sootsman, S.~N. Girard, J.-C. Zheng, J.~Wen, Y.~Zhu, M.~G.
  Kanatzidis, and V.~P. Dravid, \emph{On the Origin of Increased Phonon
  Scattering in Nanostructured PbTe Based Thermoelectric Materials}, J.\ Am.\
  Chem.\ Soc. \textbf{132}, 34--37 (2010).
\bibAnnoteFile{He-jacs-2010}

\bibitem{Chena-pnsmi-2012}
Z.-G. Chena, G.~Hana, L.~Yanga, L.~Chenga, and J.~Zoua, \emph{Nanostructured
  thermoelectric materials: Current research and future challenge}, Prog.\
  Nat.\ Sci. \textbf{22}, 535–549 (2012).
\bibAnnoteFile{Chena-pnsmi-2012}

\bibitem{Harman-jeml-2005}
T.~C. Harman, M.~P. Walsh, B.~E. Laforge, and G.~W. Turner,
  \emph{Nanostructured Thermoelectric Materials}, J.\ Elec.\ Mater.\ Lett.
  \textbf{34}, L19--L22 (2005).
\bibAnnoteFile{Harman-jeml-2005}

\bibitem{Zhao-jacs-2012}
L.-D. Zhao, J.~He, S.~Hao, C.-I. Wu, T.~P. Hogan, C.~Wolverton, V.~P. Dravid,
  and M.~G. Kanatzidis, \emph{Raising the thermoelectric performance of p-type
  PbS with endotaxial nanostructuring and valence-band offset engineering using
  CdS and ZnS.}, J.\ Am.\ Chem.\ Soc. \textbf{134}, 16327--16336 (2012).
\bibAnnoteFile{Zhao-jacs-2012}

\bibitem{zaoui_mcp_2009}
A.~Zaoui, S.~Kacimi, M.~Zaoui, and B.~Bouhafs, \emph{Theoretical investigation
  of electronic structure of PbS$_x$Te$_{1-x}$ and PbSe$_x$Te$_{1-x}$}, Mater.\
  Chem.\ Phys. \textbf{114}, 650--655 (2009).
\bibAnnoteFile{zaoui_mcp_2009}

\bibitem{naeemullaha_cms_2014}
Naeemullaha, G.~Murtazab, R.~Khenatac, N.~Hassana, S.~Naeemb, M.~N. Khalidb,
  and S.~B. Omrand, \emph{Structural and optoelectronic properties of
  PbS$_x$Se$_{1-x}$, PbS$_x$Te$_{1-x}$ and PbSe$_x$Te$_{1-x}$ first-principles
  calculations}, Comp.\ Mat.\ Sci. \textbf{83}, 496--503 (2014).
\bibAnnoteFile{naeemullaha_cms_2014}

\bibitem{yamini_pccp_2014}
S.~A. Yamini, H.~Wang, Z.~M. Gibbs, Y.~Pei, S.~X. Doua, and G.~J. Snyder,
  \emph{Chemical composition tuning in quaternary p-type Pb-chalcogenides – a
  promising strategy for enhanced thermoelectric performance}, Phys.\ Chem.\
  Chem.\ Phys. \textbf{16}, 1835--1840 (2014).
\bibAnnoteFile{yamini_pccp_2014}

\bibitem{Qzhang_jacs_2012}
Q.~Zhang, F.~Cao, W.~Liu, K.~Lukas, B.~Yu, S.~Chen, C.~Opeil, D.~Broido,
  G.~Chen, and Z.~Ren, \emph{Heavy Doping and Band Engineering by Potassium to
  Improve the Thermoelectric Figure of Merit in p-Type PbTe, PbSe, and
  PbTe$_{1–y}$Se$_y$}, J.\ Am.\ Chem.\ Soc. \textbf{134}, 10031--10038
  (2012).
\bibAnnoteFile{Qzhang_jacs_2012}

\bibitem{Pei_nature_2011}
Y.~Pei, X.~Shi, A.~LaLonde, H.~Wang, L.~Chen, and G.~J. Snyder,
  \emph{Convergence of electronic bands for high performance bulk
  thermoelectrics}, Nature \textbf{473}, 66–69 (2011).
\bibAnnoteFile{Pei_nature_2011}

\bibitem{Zhang_ees_2012}
Q.~Zhang, H.~Wang, W.~Liu, H.~Wang, B.~Yu, Q.~Zhang, Z.~Tian, G.~Ni, S.~Lee,
  K.~Esfarjani, G.~Chen, and Z.~Ren, \emph{Enhancement of thermoelectric
  figure-of-merit by resonant states of aluminium doping in lead selenide},
  Energ.\ Environ.\ Sci. \textbf{5}, 5246--5251 (2012).
\bibAnnoteFile{Zhang_ees_2012}

\bibitem{poudeu_jacs_2006}
P.~F.~P. Poudeu, J.~D’Angelo, H.~Kong, A.~Downey, J.~L. Short, R.~Pcionek,
  T.~P. Hogan, C.~Uher, and M.~G. Kanatzidis, \emph{Nanostructures versus Solid
  Solutions: Low Lattice Thermal Conductivity and Enhanced Thermoelectric
  Figure of Merit in Pb$_{9.6}$Sb$_{0.2}$Te$_{10-x}$Se$_x$ Bulk Materials}, J.\
  Am.\ Chem.\ Soc. \textbf{128}, 14347--14355 (2006).
\bibAnnoteFile{poudeu_jacs_2006}

\bibitem{Androulakis_jacs_2015}
J.~Androulakis, C.-H. Lin, H.-J. Kong, C.~Uher, C.-I. Wu, T.~Hogan, B.~A. Cook,
  T.~Caillat, K.~M. Paraskevopoulos, and M.~G. Kanatzidis, \emph{Spinodal
  Decomposition and Nucleation and Growth as a Means to Bulk Nanostructured
  Thermoelectrics: Enhanced Performance in Pb$_{1-x}$Sn$_x$Te-PbS}, J.\ Am.\
  Chem.\ Soc. \textbf{129}, 9780--9788 (2007).
\bibAnnoteFile{Androulakis_jacs_2015}

\bibitem{boukhris_ps_2011}
N.~Boukhris, H.~Meradji, S.~Ghemid, S.~Drablia, and F.~E.~H. Hassan, \emph{{\it
  Ab initio} study of the structural, electronic and thermodynamic properties
  of PbSe$_{1-x}$S$_x$, PbSe$_{1-x}$Te$_x$ and PbS$_{1-x}$Te$_x$ ternary
  alloys}, Phys.\ Scripta \textbf{83}, 065701--065710 (2011).
\bibAnnoteFile{boukhris_ps_2011}

\bibitem{Liu_MineMag_1994}
H.~Liu and L.~L.~Y. Chang, \emph{Phase relation in the system PbS-PbSe-PbTe},
  Mineral.\ Mag. \textbf{58}, 567--578 (1994).
\bibAnnoteFile{Liu_MineMag_1994}

\bibitem{curtarolo:art49}
O.~Levy, G.~L.~W. Hart, and S.~Curtarolo, \emph{Uncovering compounds by synergy
  of cluster expansion and high-throughput methods}, J.\ Am.\ Chem.\ Soc.
  \textbf{132}, 4830--4833 (2010).
\bibAnnoteFile{curtarolo:art49}

\bibitem{curtarolo:art53}
O.~Levy, R.~V. Chepulskii, G.~L.~W. Hart, and S.~Curtarolo, \emph{The New face
  of {Rhodium} alloys: revealing ordered structures from first principles}, J.\
  Am.\ Chem.\ Soc. \textbf{132}, 833--837 (2010).
\bibAnnoteFile{curtarolo:art53}

\bibitem{curtarolo:art56}
R.~H. Taylor, S.~Curtarolo, and G.~L.~W. Hart, \emph{Predictions of the
  {Pt$_8$Ti} phase in unexpected systems}, J.\ Am.\ Chem.\ Soc. \textbf{132},
  6851--6854 (2010).
\bibAnnoteFile{curtarolo:art56}

\bibitem{saunders_1998}
N.~Saunders and A.~P. Miodownik, \emph{CALPHAD: A Comprehensive Guide},
  Pergamon Materials Series (Elsevier Science, UK, 1998).
\bibAnnoteFile{saunders_1998}

\bibitem{hillert_2008}
M.~Hillert, \emph{Phase Equilibria, Phase Diagrams and Phase Transformations}
  (Cambridge University Press, UK, 2008).
\bibAnnoteFile{hillert_2008}

\bibitem{Barin}
I.~Barin, \emph{Thermochemical Data of pure substances} (VCH, Germany, 1993).
\bibAnnoteFile{Barin}

\bibitem{Zhang_ActaMat_2007}
R.~F. Zhang and S.~Verpek, \emph{Phase stabilities and spinodal decomposition
  in the Cr$_{1-x}$Al$_x$N system strudied by ab initio LDA and
  thermodynamicmodelling: Comparision with the Ti$_{1-x}$Al$_x$N and
  TiN/Si$_{3}$N$_4$ systems}, Acta\ Mater. \textbf{55}, 4615--4624 (2007).
\bibAnnoteFile{Zhang_ActaMat_2007}

\bibitem{Sheng_ActaMat_2011}
S.~H. Sheng, R.~F. Zhang, and S.~Verpek, \emph{Phase stabilities and
  decomposition mechanism in the Zr–Si–N system studied by combined ab
  initio DFT and thermodynamic calculation}, Acta\ Mater. \textbf{59}, 297--307
  (2011).
\bibAnnoteFile{Sheng_ActaMat_2011}

\bibitem{SHSheng_ActaMat_2011}
S.~H. Sheng, R.~F. Zhang, and S.~Verpek, \emph{Study of spinodal decomposition
  and formation of nc-Al$_2$O$_3$/ZrO$_2$ nanocomposites by combined ab initio
  density functional theory and thermodynamic modeling}, Acta\ Mater.
  \textbf{59}, 3498--3509 (2011).
\bibAnnoteFile{SHSheng_ActaMat_2011}

\bibitem{Redlich_IndsEngChem_1948}
O.~Redlich and A.~Kister, \emph{Algebraic representation of Thermodynamic
  properties and the classification of solutions}, Ind.\ Eng.\ Chem.\ Res.
  \textbf{40}, 345--348 (1948).
\bibAnnoteFile{Redlich_IndsEngChem_1948}

\bibitem{curtarolo:art81}
S.~Curtarolo, G.~L.~W. Hart, M.~{Buongiorno~Nardelli}, N.~Mingo, S.~Sanvito,
  and O.~Levy, \emph{The high-throughput highway to computational materials
  design}, Nat.\ Mater. \textbf{12}, 191--201 (2013).
\bibAnnoteFile{curtarolo:art81}

\bibitem{aflowBZ}
W.~Setyawan and S.~Curtarolo, \emph{High-throughput electronic band structure
  calculations: challenges and tools}, Comp.\ Mat.\ Sci. \textbf{49}, 299--312
  (2010).
\bibAnnoteFile{aflowBZ}

\bibitem{aflowlibPAPER}
S.~Curtarolo, W.~Setyawan, S.~Wang, J.~Xue, K.~Yang, R.~H. Taylor, L.~J.
  Nelson, G.~L.~W. Hart, S.~Sanvito, M.~{Buongiorno~Nardelli}, N.~Mingo, and
  O.~Levy, \emph{AFLOWLIB.ORG: A distributed materials properties repository
  from high-throughput {\it ab initio} calculations}, Comp.\ Mat.\ Sci.
  \textbf{58}, 227--235 (2012).
\bibAnnoteFile{aflowlibPAPER}

\bibitem{deFontaine_ssp_1994}
D.~de~Fontaine, \emph{Cluster Approach to Order-Disorder Transformations in
  Alloys}, in \emph{Solid State Physics}, edited by H.~Ehrenreich and
  D.~Turnbull (Wiley, New York, 1994), vol.~47, pp. 33--176.
\bibAnnoteFile{deFontaine_ssp_1994}

\bibitem{burton_cg_2006}
B.~P. Burton and A.~van~de Walle, \emph{First-principle phase diagram
  calculations for the system NaCl-KCl: The role of excess vibrational
  entropy}, Chem.\ Geol. \textbf{225}, 222--229 (2006).
\bibAnnoteFile{burton_cg_2006}

\bibitem{Walle_calphad_2002}
A.~van~de Walle, M.~Asta, and G.~Ceder, \emph{The alloy theoretic automated
  toolkit: A user guide}, Calphad \textbf{26}, 539--553 (2002).
\bibAnnoteFile{Walle_calphad_2002}

\bibitem{Walle_jpe_2002}
A.~van~de Walle and G.~Ceder, \emph{Automating First-Principles Phase Diagram
  Calculations}, J.\ Phase Equilib. \textbf{23}, 348--359 (2002).
\bibAnnoteFile{Walle_jpe_2002}

\bibitem{Walle_msmse_2002}
A.~van~de Walle and M.~Asta, \emph{Self-driven lattice-model Monte Carlo
  simulations of alloy thermodynamic properties and phase diagrams}, Model.\
  Simul.\ Mater.\ Sc. \textbf{10}, 521--538 (2002).
\bibAnnoteFile{Walle_msmse_2002}

\bibitem{curtarolo:art65}
S.~Curtarolo, W.~Setyawan, G.~L.~W. Hart, M.~Jahnatek, R.~V. Chepulskii, R.~H.
  Taylor, S.~Wang, J.~Xue, K.~Yang, O.~Levy, M.~Mehl, H.~T. Stokes, D.~O.
  Demchenko, and D.~Morgan, \emph{AFLOW: an automatic framework for
  high-throughput materials discovery}, Comp.\ Mat.\ Sci. \textbf{58}, 218--226
  (2012).
\bibAnnoteFile{curtarolo:art65}

\bibitem{curtarolo:art92}
R.~H. Taylor, F.~Rose, C.~Toher, O.~Levy, K.~Yang, M.~{Buongiorno~Nardelli},
  and S.~Curtarolo, \emph{A RESTful API for exchanging Materials Data in the
  AFLOWLIB.org consortium}, Comp. Mat. Sci. \textbf{93}, 178--192 (2014).
\bibAnnoteFile{curtarolo:art92}

\bibitem{kresse_vasp}
G.~Kresse and J.~Hafner, \emph{{\it Ab initio} molecular dynamics for liquid
  metals}, Phys.\ Rev.\ B \textbf{47}, 558--561 (1993).
\bibAnnoteFile{kresse_vasp}

\bibitem{calderon_2015}
C.~E. Calderon, J.~J. Plata, C.~Toher, C.~Oses, O.~Levy, M.~Fornari, A.~Natan,
  M.~J. Mehl, G.~Hurt, M.~B. Nardelli, and S.~Curtarolo, \emph{The AFLOW
  Standard for High-Throughput Materials Science Calculations diagrams}, Comp.\
  Mat.\ Sci. \textbf{108 Part A}, 233--238 (2015).
\bibAnnoteFile{calderon_2015}

\bibitem{PAW}
P.~E. Bl\"ochl, \emph{Projector augmented-wave method}, Phys.\ Rev.\ B
  \textbf{50}, 17953--17979 (1994).
\bibAnnoteFile{PAW}

\bibitem{PBE}
J.~P. Perdew, K.~Burke, and M.~Ernzerhof, \emph{Generalized gradient
  approximation made simple}, Phys.\ Rev.\ Lett. \textbf{77}, 3865--3868
  (1996).
\bibAnnoteFile{PBE}

\bibitem{cottrell1967}
A.~Cottrell, \emph{An introduction to metallurgy} (St. Martin's Press, UK,
  1967).
\bibAnnoteFile{cottrell1967}

\bibitem{Ferreira_PRB_2006}
L.~G. Ferreira, M.~Marques, and L.~K. Teles, \emph{Ga$_{1-x}$Al$_x$N system,
  Madelung, and strain energies: A strudy on the quality of cluster expansion},
  Phys.\ Rev.\ B \textbf{74}, 075324--075324 (2006).
\bibAnnoteFile{Ferreira_PRB_2006}

\bibitem{Marques_PRB_2007}
M.~Marques, L.~K. Teles, and L.~G. Ferreira, \emph{Influence of miscibility on
  the energy-gap dispersion in Al$_x$Ga$_{1-x}$N alloys: First-principles
  calculations}, Phys.\ Rev.\ B \textbf{75}, 033201 (2007).
\bibAnnoteFile{Marques_PRB_2007}

\bibitem{Darrow_JMatSci_1969}
M.~S. Darrow, W.~B. White, and R.~Roy, \emph{Micro-Indentation Hardness
  Variation as a Function of Composition for Polycrystalline Solution in the
  Systems PbS/PbTe, PnSe/PbTe, and PbS/PbSe}, J.\ Mater.\ Sci. \textbf{4},
  313--319 (1969).
\bibAnnoteFile{Darrow_JMatSci_1969}

\bibitem{burton_jap_2011}
B.~P. Burton, S.~Demers, and A.~van~de Walle, \emph{First-principle phase
  diagram calculations for the wurtzite-structure quasibinary systems SiC-AlN,
  SiC-GaN and SiC-InN}, J.\ Appl.\ Phys. \textbf{110}, 023507 (2011).
\bibAnnoteFile{burton_jap_2011}

\end{thebibliography}
\end{document}